\documentclass[11pt]{article}
\usepackage{amsmath}
\usepackage{amsfonts}
\usepackage{amssymb}
\usepackage{graphicx}
\usepackage{bm}
\usepackage{color}
\usepackage{amscd}

\def\bea{\begin{eqnarray}}
\def\eea{\end{eqnarray}}

\begin{document}
\begin{center}
\LARGE {\bf The Heisenberg algebra as near horizon symmetry of the black flower solutions of Chern-Simons-like theories of gravity}
\end{center}

\begin{center}
{M. R. Setare \footnote{E-mail: rezakord@ipm.ir}\hspace{1mm} ,
H. Adami \footnote{E-mail: hamed.adami@yahoo.com}\hspace{1.5mm} \\
{\small {\em  Department of Science, University of Kurdistan, Sanandaj, Iran.}}}\\

\end{center}

\begin{center}
{\bf{Abstract}}\\
In this paper we study the near horizon symmetry algebra of the non-extremal black hole solutions of the Chern-Simons-like theories of gravity, which are stationary but are not necessarily spherically symmetric. We define the extended off-shell ADT current which is an extension of the generalized ADT current. We use the extended off-shell ADT current to define quasi-local conserved charges such that they are conserved for Killing vectors and asymptotically Killing vectors which depend on dynamical fields of the considered theory. We apply this formalism to the Generalized Minimal Massive Gravity( GMMG) and obtain conserved charges of a spacetime which describes near horizon geometry of non-extremal black holes. Eventually, we find the algebra of conserved charges in Fourier modes. It is interesting that, similar to the Einstein gravity in the presence of negative cosmological constant, for the GMMG model also we obtain the Heisenberg algebra as the near horizon symmetry algebra of the black flower solutions. Also the vacuum state and all descendants of the vacuum have the same energy. Thus these zero energy excitations on the horizon appear as soft hairs on the black hole.
\end{center}

\section{Introduction}
It is well known that the pure Einstein-–Hilbert gravity in three dimensions exhibits no propagating physical degrees of freedom  \cite{2',3'}. So choosing appropriate conditions at
the boundary is crucial in this theory. Depending on the chosen boundary conditions, this theory can lead to
completely different boundary theories. Adding the gravitational Chern-Simons term produces a
propagating massive graviton \cite{1'}. The resulting
theory is called topologically massive gravity (TMG). Including a
negative cosmological constant, yields cosmological topologically massive
gravity (CTMG). In this case the theory exhibits both gravitons and black
holes. Unfortunately there is a problem in this model, with the usual sign for
the gravitational constant, the massive excitations of CTMG carry negative
energy. In the absence of a cosmological constant, one can change the sign
of the gravitational constant, but if $\Lambda<0$, this will give a negative mass to
the BTZ black hole, so the existence of a stable ground state is in doubt in
this model \cite{1000,1008}. A few years ego a new theory of massive gravity (NMG) in three dimensions
has been proposed \cite{2''}. This theory is equivalent to the three-dimensional
Fierz-Pauli action for a massive spin-2 field at the linearized level. With the only Einstein- Hilbert term in the action there are no propagating degrees of freedom, but by adding the higher curvature terms in the action the situation becomes different.
Usually the theories including the terms given by the square of the curvatures have the massive spin
2 mode and the massive scalar mode in addition to the massless graviton. Also the theory has ghosts
due to negative energy excitations of the massive tensor. The unitarity of NMG was discussed in \cite{1001,1002} (see also \cite{1003,1004}) and this model is generalized to higher dimensions. It was found that
there exist a choice of parameters for which these theories possess one AdS background on which
neither massive fields, nor massless scalars propagate. By this special choice of the parameters,
which is called as a critical point, there appears a mode which behaves as a logarithmic function
of the distance. The massive graviton modes obey Brown-Henneaux boundary conditions, at the
critical point in parameter space, the massive gravitons become massless and are replaced by
new modes, so-called the logarithmic modes. Although, it has been shown the compliance of the
NMG with the holographic c-theorem \cite{1005,1006}, both TMG and NMG have
a bulk-boundary unitarity conflict. In another term either the bulk or the
boundary theory is non-unitary, so there is a clash between the positivity
of the two Brown-Henneaux boundary c charges and the bulk energies \cite{1007}. Recently an
interesting three dimensional massive gravity introduced by Bergshoeff, et.
al \cite{3''} which dubbed Minimal Massive Gravity (MMG), which has the same
minimal local structure as TMG. The MMG model has the same gravitational
degree of freedom as the TMG has and the linearization of the metric
field equations for MMG yield a single propagating
massive spin-2 field. It seems that the single massive degree of freedom of MMG is
unitary in the bulk and gives rise to a unitary CFT on the boundary. More recently the author of \cite{5'} has introduced  Generalized Minimal Massive Gravity (GMMG), an interesting modification of MMG. GMMG is a unification of MMG with NMG, so this model is realized by adding higher-derivative deformation term to the Lagrangian of MMG.
As has been shown in \cite{5'}, GMMG also avoids the aforementioned ``bulk-boundary unitarity clash''.
Calculation of the GMMG action to quadratic order about $AdS_3$ space show that the theory is free of negative-energy bulk modes. Also Hamiltonian analysis show that the GMMG model has no Boulware-Deser ghosts and this model propagate only two physical modes.
 So these models are viable candidate for semi-classical limit of a unitary quantum $3D$ massive gravity.\\
 Although the Chern-Simons-like theories of gravity (CSLTG) in $(2+1)$-dimension \cite{1} (e.g. TMG , NMG, MMG , Zwei-dreibein gravity (ZDG)\cite{4'}, GMMG, etc), exhibit local physical degrees
of freedom, but for these theories also, different boundary conditions can lead to
completely different boundary theories. For matter-free Einstein-–Hilbert gravity, the behavior of the three-dimensional metric at spatial infinity
is given by the Brown-Henneaux boundary conditions \cite{12}. But in the
presence of matter, these boundary conditions can be modified \cite{13}. This modification can be occurs in Topological massive gravity \cite{14} and even in pure Einstein-–Hilbert gravity \cite{15}.\\
Recently the authors of \cite{9} have considered the black flower solution of the Einstein
equations in $3d$ \cite{16}, then have proposed
a new set of boundary conditions, which leads to a very simple near horizon symmetry algebra,
the Heisenberg algebra. \footnote{Here we should mention that  Donnay et al \cite{500}, have shown that the asymptotic symmetries close to the horizon of the nonextremal black hole solution of the three-dimensional Einstein gravity in the presence of
a negative cosmological term, are generated
by an extension of supertranslations (see also \cite{140}). The near
horizon symmetries in three dimensions are related with the Bondi–-van der Burg–-Metzner–-Sachs (BMS) algebra \cite{500'}. The authors of \cite{500} have shown that for a special choice of boundary conditions, the
near region to the horizon of a stationary black hole
present a generalization of supertranslation, including a
semidirect sum with superrotations, represented by Virasoro
algebra. From BMS supertranslation we know that the vacuum is not
unique and infinite degenerate vacua are physically distinct and are related to each
other by the BMS supertranslation (for more information see \cite{501,502,503}).} In this paper we are going to study this near horizon symmetry in the framework of Chern-Simons-like theories of gravity. For this purpose, at first we should obtain boundary conserved charges. Here we use a formalism  based on the concept of quasi-local conserved charges.
 We have obtained  the quasi-local conserved charges of the Lorentz-diffeomorphism covariant theories of gravity in the first order formalism, in paper \cite{29}. In previous paper \cite{150} by introducing the total variation of a quantity due to the infinitesimal Lorentz-diffeomorphism transformation, we have obtained the conserved charges in the Lorentz-diffeomorphism non-covariant
theories. Here we should find an expression for the quasi-local conserved charges of CSLTG associated with the field dependent Killing vector fields. So we need an extended version of the generalized off-shell ADT current such that it becomes conserved for the field dependent Killing vectors and the field dependent asymptotically Killing vectors. By this extension, we obtain the quasi-local conserved charge corresponds to a field dependent Killing vector field. After that we apply our formalism to the  GMMG and obtain conserved charges of the non-extremal black hole solutions  which are stationary but are not necessarily spherically symmetric. By writing conserved charges in Fourier modes, we find the Heisenberg algebra as the near horizon symmetry algebra of these black solutions. Similar to the Einstein gravity \cite{9}, for the GMMG also, we obtain the Hamiltonian as  $H \equiv P_{0}$, where  $P_{0}$ is a Casimir of the algebra, so the vacuum state and all descendants of the vacuum have the same energy. These zero energy excitations on the horizon appear as soft hairs on the black hole. By setting $\sigma =-1$, $\mu \rightarrow \infty$ and $m^{2} \rightarrow \infty$, where the GMMG reduce to the Einstein gravity with negative cosmological constant, all our results for the GMMG, reduced to the results of \cite{9} which have been obtained by a different way.
\section{Quasi-local conserved charges associated with field dependent Killing vectors}
In this section we consider Chern-Simons-like theories of gravity, then we find an expression for quasi-local conserved charge corresponds to a field dependent Killing vector which is admitted by a solution of considered theory. The Lagrangian 3-form of the CSLTG is given by \cite{1}
\begin{equation}\label{1}
  L=\frac{1}{2} \tilde{g}_{rs} a^{r} \cdot da^{s}+\frac{1}{6} \tilde{f}_{rst} a^{r} \cdot a^{s} \times a^{t}.
\end{equation}
In the above Lagrangian $ a^{ra}=a^{ra}_{\hspace{3 mm} \mu} dx^{\mu} $ are the Lorentz vector valued one-forms, where $r=1,...,N$ and $a$ indices refer to the flavour and the Lorentz indices, respectively. We should mention that, here the wedge products of the Lorentz-vector valued one-form fields are implicit. Also, $\tilde{g}_{rs}$ is a symmetric constant metric on the flavour space and $\tilde{f}_{rst}$ are the totally symmetric "flavour tensors" which are interpreted as the coupling constants. We use a 3D-vector algebra notation for the Lorentz vectors in which contractions with $\eta _{ab}$ and $\varepsilon ^{abc}$ are denoted by dots and crosses, respectively \footnote{Here we consider the notation used in \cite{1}.}. It is worth saying that $a^{ra}$ is a collection of the dreibein $e^{a}$, the dualized spin-connection $\omega ^{a}$, the auxiliary field $ h^{a}_{\hspace{1.5 mm} \mu} = e^{a}_{\hspace{1.5 mm} \nu} h^{\nu}_{\hspace{1.5 mm} \mu} $ and so on. Also for all interesting CSLTG we have $\tilde{f}_{\omega rs} = \tilde{g}_{rs}$ \cite{2} \footnote{The Lagrangian of CSLTG contains combinations such as $f \cdot R=f \cdot d \omega + \frac{1}{2}f \cdot \omega \times \omega$, $f \cdot D(\omega) h = f \cdot d h + \omega \cdot f \times h$, $\omega \cdot d \omega + \frac{1}{2} \omega \cdot \omega \times \omega$ and so on. It can be seen that all of these combinations obey the equation $\tilde{f}_{\omega rs} = \tilde{g}_{rs}$.}.\\
Let $\pounds_{\xi}$ denotes the ordinary Lie derivative along $\xi$ and the Lie-Lorentz derivative (L-L derivative) $\mathfrak{L}_{\xi}$ is defined by \cite{3}
\begin{equation}\label{2}
  \mathfrak{L}_{\xi} A^{a \cdots } _{ \hspace{4.5 mm} b \cdots } = \pounds_{\xi} A^{a \cdots } _{ \hspace{4.5 mm} b \cdots } + \lambda ^{a}_{ \xi \hspace{1 mm} c} A^{c \cdots } _{ \hspace{4.5 mm} b \cdots } + \cdots - \lambda ^{c}_{\xi \hspace{1 mm} b} A^{a \cdots } _{ \hspace{4.5 mm} c \cdots } - \cdots ,
\end{equation}
where $\lambda ^{ab}_{\xi}$ is generator of the Lorentz gauge transformations $SO(2,1)$. The total variation of $a^{ra}$ due to a diffeomorphism generator $\xi$ is \cite{4}
\begin{equation}\label{3}
  \delta _{\xi} a^{ra} = \mathfrak{L}_{\xi} a^{ra} -\delta ^{r} _{\omega} d \chi _{\xi} ^{a} ,
\end{equation}
which is caused by a combination of variations due to the diffeomorphism and the infinitesimal Lorentz gauge transformation. In Eq.\eqref{3}, $\chi _{\xi} ^{a}= \frac{1}{2} \varepsilon ^{a} _{\hspace{1.5 mm} bc} \lambda _{\xi}^{bc} $ and $ \delta ^{r} _{s} $  denotes the ordinary Kronecker delta. Also, $\chi _{\xi} ^{a}$ is a general function of space-time coordinates and of the diffeomorphism generator $\xi$. It should be noted that $\chi _{\xi} ^{a}$ is linear in $\xi$. One can find the total variation of the Lagrangian due to the diffeomorphism generator $ \xi $ as \cite{4}
\begin{equation}\label{4}
  \delta _{\xi} L = \mathfrak{L}_{\xi} L +d \psi _{\xi} ,
\end{equation}
where $\psi _{\xi}$ is given by
\begin{equation}\label{5}
   \psi _{\xi} = \frac{1}{2} \tilde{g}_{\omega r} d \chi _{\xi} \cdot a^{r} .
\end{equation}
The variation of the Lagrangian \eqref{1} is given by
\begin{equation}\label{6}
  \delta L = \delta a^{r} \cdot E_{r} + d \Theta (a, \delta a) ,
\end{equation}
where
\begin{equation}\label{7}
   E_{r}^{\hspace{1.5 mm} a} = \tilde{g}_{rs} d a^{sa} + \frac{1}{2} \tilde{f}_{rst} (a^{s} \times a^{t})^{a} ,
\end{equation}
so that $E_{r}^{\hspace{1.5 mm} a}=0$ are the equations of motion, and
\begin{equation}\label{8}
   \Theta (a, \delta a) = \frac{1}{2} \tilde{g}_{rs} \delta a^{r} \cdot a^{s},
\end{equation}
is the surface term. The total variation of the surface term is
\begin{equation}\label{9}
\delta _{\xi} \Theta (a , \delta a ) = \mathfrak{L}_{\xi} \Theta (a , \delta a ) + \Pi _{\xi},
\end{equation}
where
\begin{equation}\label{10}
\Pi _{\xi}= \frac{1}{2} g_{\omega r} d \chi _{\xi} \cdot \delta a^{r} .
\end{equation}
Now, by considering that the variation in Eq.\eqref{6} is the total variation generated by $\xi$ and by using the Bianchi identities, we find that \cite{4}
\begin{equation}\label{11}
d J_{\xi} = 0,
\end{equation}
where
\begin{equation}\label{12}
  J_{\xi} = \Theta (a, \delta _{\xi} a ) - i_{ \xi } L - \psi _{\xi} + i_{\xi} a^{r} \cdot E _{r} - \chi _{\xi} \cdot E _{\omega},
\end{equation}
 here $i_{\xi}$ denotes interior product in $\xi$. Strictly speaking, $J_{\xi}$ is an off-shell conserved current , i.e. the equation \eqref{11} is hold off-shell. By virtue of the Poincare lemma, one can write $J_{\xi}=d K_{\xi}$. It is easy to show that
\begin{equation}\label{13}
    K_{\xi} = \frac{1}{2} \tilde{g}_{rs} i_{\xi} a^{r} \cdot a^{s} - \tilde{g} _{\omega s} \chi _{\xi} \cdot a^{s} ,
\end{equation}
Let $\hat{\delta}$ denotes variation due to dynamical fields. By varying Eq.\eqref{12} with respect to dynamical fields we will have
\begin{equation}\label{14}
  \begin{split}
     d \left( \hat{\delta} K_{\xi} - K_{\hat{\delta}\xi} - i_{\xi} \Theta (a, \hat{\delta} a) \right) = & \hat{\delta} \Theta (a, \delta _{\xi} a) - \delta _{\xi} \Theta (a, \hat{\delta} a) - \Theta (a, \delta _{\hat{\delta} \xi} a) \\
       & + \hat{\delta} a^{r} \cdot i_{\xi} E_{r} + i_{\xi} a^{r} \cdot \hat{\delta} E_{r} - \chi _{\xi} \cdot \hat{\delta} E_{\omega}.
  \end{split}
\end{equation}
In the calculation of the above equation, we assumed that $\xi$ is a function of dynamical fields and we used the fact that $\hat{\delta} \chi _{\xi} = \chi _{\hat{\delta} \xi}$, because $\chi _{\xi}$ is linear in $\xi$. We define the right hand side of Eq.\eqref{14} as extended off-shell ADT current, namely
\begin{equation}\label{15}
\begin{split}
   \mathfrak{J}_{ADT} (a , \hat{\delta} a, \delta _{\xi} a)=  & \hat{\delta} a^{r} \cdot i_{\xi} E_{r} + i_{\xi} a^{r} \cdot \hat{\delta} E_{r} - \chi _{\xi} \cdot \hat{\delta} E_{\omega} \\
     & + \hat{\delta} \Theta (a, \delta _{\xi} a) - \delta _{\xi} \Theta (a, \hat{\delta} a) - \Theta (a, \delta _{\hat{\delta} \xi} a).
\end{split}
\end{equation}
The extended off-shell ADT current will be reduced to the generalized off-shell ADT current \cite{4} when $\xi$ is independent of the dynamical fields, that is $\hat{\delta} \xi =0$.
The extended off-shell ADT current $\mathfrak{J}_{ADT}$ reduces to
\begin{equation}\label{16}
  \Omega _{Symp}(a , \hat{\delta} a, \delta _{\xi} a)= \hat{\delta} \Theta (a, \delta _{\xi} a) - \delta _{\xi} \Theta (a, \hat{\delta} a) - \Theta (a, \delta _{\hat{\delta} \xi} a),
\end{equation}
when the equations of motion and the linearized equations of motion both are satisfied. The equation \eqref{16} is just the ordinary symplectic current \cite{5,6,7,8} when $\xi$ is independent of dynamical fields, that is $\hat{\delta} \xi =0$. So it seems sensible that Eq.\eqref{16} is an extension of the symplectic current. By substituting Eq.\eqref{8} into  Eq.\eqref{16} we have
\begin{equation}\label{17}
  \Omega _{Symp}(a , \hat{\delta} a, \delta _{\xi} a)= \tilde{g} _{rs} \delta _{\xi} a^{r} \cdot \hat{\delta} a^{s}.
\end{equation}
By replacing $\hat{\delta} = \delta_{1}$ and $\delta _{\xi} = \delta_{2}$, the equation \eqref{17} becomes
\begin{equation}\label{18}
  \Omega _{Symp}(a , \delta_{1} a, \delta_{2} a)= \tilde{g} _{rs} \delta_{2} a^{r} \cdot \delta_{1} a^{s}.
\end{equation}
It is clear that $\Omega _{Symp}$ is closed, skew-symmetric and non-degenerate, also it explicitly vanishes when $\xi$ is a Killing vector field, namely $\delta _{\xi} a^{r} =0$, then $\Omega _{Symp}$ has all properties of a symplectic current.\\
On the other hand, in the off-shell case, if we assume that $\xi$ is a Killing vector field, then
\begin{equation}\label{19}
   \hat{\delta} \Theta (a, \delta _{\xi} a) - \delta _{\xi} \Theta (a, \hat{\delta} a) - \Theta (a, \delta _{\hat{\delta} \xi} a) =0.
\end{equation}
Thus, in this case, the extended off-shell ADT current reduces to the ordinary one \cite{4}. So, by the above discussion, the definition of the extended off-shell ADT current as Eq.\eqref{15} makes sense.\\
Now we can write Eq.\eqref{14} as follows:
\begin{equation}\label{20}
  \mathfrak{J}_{ADT} (a , \hat{\delta} a, \delta _{\xi} a) = d \mathfrak{Q}_{ADT} (a , \hat{\delta} a;\xi),
\end{equation}
where $\mathfrak{Q}_{ADT}$ is extended off-shell ADT conserved charge and it is defined as
\begin{equation}\label{21}
  \mathfrak{Q}_{ADT} (a , \hat{\delta} a;\xi) = \hat{\delta} K_{\xi} - K_{\hat{\delta}\xi} - i_{\xi} \Theta (a, \hat{\delta} a).
\end{equation}
It should be noted that, the first term in the right hand side of the above equation, is just the Komar expression for the charge perturbation \cite{26}. Second term comes from the fact that, it is assumed that $\xi$ is dependent on the dynamical fields and
the third term is the contribution of surface term in charge perturbation  \cite{5,6,7,8}. In this way, we can define quasi-local conserved charge perturbation associated with a field dependent vector field $\xi$ as
\begin{equation}\label{22}
 \hat{\delta} Q(\xi) = \frac{1}{8 \pi G} \int _{\Sigma} \mathfrak{Q}_{ADT} (a , \hat{\delta} a;\xi),
\end{equation}
where $G$ denotes the Newtonian gravitational constant and $\Sigma$ is a space-like codimension two surface. Due to the definition \eqref{15}, the quasi-local conserved charge \eqref{22} is not only conserved for Killing vectors which are admitted by spacetime everywhere but also it is conserved for the asymptotic Killing vectors. By substituting Eq.\eqref{8} and Eq.\eqref{13} into Eq.\eqref{22} we find that
\begin{equation}\label{23}
 Q ( \xi )  = \frac{1}{8 \pi G} \int_{0}^{1} ds \int_{\Sigma} \left( g_{rs} i_{\xi} a^{r} - g _{\omega s} \chi _{\xi} \right) \cdot \hat{\delta} a^{s},
\end{equation}
where we took an integration from \eqref{22} over the one-parameter path on the solution space \cite{10,11,4}. This has exactly the form of case in which the Killing vector field is independent of dynamical fields \cite{4}. However, we argued that it is usable for the case in which $\xi$ is field dependent.\\
The symplectic current \eqref{16} vanishes when $\xi$ is a Killing vector field\footnote{ In this paragraph, we drop "field dependent" phrase for simplicity.} admitted by spacetime everywhere, then it is easy to see from \eqref{14} that, the generalized off-shell ADT current becomes conserved for this case. However, if we assume that $\xi$ to be an asymptotically Killing vector field, the generalized off-shell ADT current is no longer a conserved quantity, instead the extended off-shell ADT current \eqref{15} is a conserved current (see Eq.\eqref{20}). Since
we have $\delta_{\xi}a^{r}= 0$ asymptotically, then the symplectic current vanishes asymptotically. Hence, the extended off-shell ADT current asymptotically reduces to the generalized off-shell ADT current. Therefore the extended off-shell ADT current is appropriate to obtain conserved charges associated with asymptotically Killing vectors.
\section{Extended near horizon geometry}
In the paper \cite{9}, the authors have proposed following metric as a new fall-off condition for near horizon of a non-extremal black hole in three dimension
\begin{equation}\label{24}
\begin{split}
    ds^{2}= & \left[ l \rho \left( f_{+} \zeta ^{+} + f_{-} \zeta ^{-} \right) + \frac{l^{2}}{4} \left( \zeta ^{+} - \zeta ^{-} \right)^{2} \right] dv ^{2} + 2 l dv d\rho \\
     & + l \left( \frac{\mathcal{J}^{+}}{\zeta ^{+}} - \frac{\mathcal{J}^{-}}{\zeta ^{-}} \right) d \rho d \phi + l \rho \left( \frac{\mathcal{J}^{+}}{\zeta ^{+}} - \frac{\mathcal{J}^{-}}{\zeta ^{-}} \right) \left( f_{+} \zeta ^{+} + f_{-} \zeta ^{-} \right) dv d \phi \\
     & + \left[ \frac{l^{2}}{4} \left( \mathcal{J}^{+} + \mathcal{J}^{-}\right)^{2} - \frac{l \rho}{\zeta ^{+}\zeta ^{-}} \left( f_{+} \zeta ^{+} + f_{-} \zeta ^{-} \right) \mathcal{J}^{+} \mathcal{J}^{-} \right] d \phi ^{2},
\end{split}
\end{equation}
where $l$ is AdS radii, $\zeta ^{\pm}$ are constant parameters, $\mathcal{J}^{\pm} = \mathcal{J}^{\pm}(\phi)$ are arbitrary functions of $\phi$ and $ f _{\pm} = f_{\pm}(\rho)$ are given as
\begin{equation}\label{25}
  f_{\pm}(\rho) = 1-\frac{\rho}{2l\zeta ^{\pm}}.
\end{equation}
The line-element \eqref{24} is written in ingoing Eddington-Finkelstein coordinates, also $v$, $\rho$ and $\phi$ are the advanced time, the radial coordinate and the angular coordinate, respectively. In the particular case of $\zeta ^{\pm}=-a$, where the constant $a$ is the Rindler acceleration, the line-element \eqref{24} will be reduced to
\begin{equation}\label{26}
\begin{split}
   ds^{2}= & - 2 a l \rho f(\rho) dv^{2} + 2 l dv d\rho -2 a^{-1} \theta(\phi) d \phi d \rho + 4 \rho \theta(\phi) f(\rho) d v d \phi \\
     & + \left[ \gamma (\phi) ^{2} + \frac{2 \rho}{a l} f(\rho) \left(\gamma (\phi) ^{2} - \theta(\phi)^{2}\right) \right] d \phi ^{2},
\end{split}
\end{equation}
where $l \mathcal{J}^{\pm} =  \gamma \pm \theta$ and $f(\rho)=1+ \frac{\rho}{2la}$. The line-element \eqref{26} describes a spacetime which possesses an event horizon located at $\rho=0$. The line-element \eqref{24} solves the Einstein equations with negative cosmological constant
\begin{equation}\label{27}
  R(\Omega)+\frac{1}{2l^{2}}e \times e =0 , \hspace{1 cm} T(\Omega)=0,
\end{equation}
where $R(\Omega) = d \Omega + \frac{1}{2} \Omega \times \Omega$ is curvature 2-form, $T(\Omega)= D(\Omega)e$ is torsion 2-form and $\Omega$ is torsion free spin-connection. Also, $D(\Omega)$ denotes exterior covariant derivative with respect to $\Omega$.\\
The following Killing vector
\begin{equation}\label{28}
  \begin{split}
       & \xi ^{v} = \frac{1}{2} \left\{ - \left( \frac{1}{\zeta ^{+}} - \frac{1}{\zeta ^{-}}\right) \left( \frac{\mathcal{J}^{+}}{\zeta ^{+}} - \frac{\mathcal{J}^{-}}{\zeta ^{-}} \right) \left( \frac{\mathcal{J}^{+}}{\zeta ^{+}} + \frac{\mathcal{J}^{-}}{\zeta ^{-}} \right)^{-1} + \left( \frac{1}{\zeta ^{+}} + \frac{1}{\zeta ^{-}}\right) \right\} \Xi (\phi) \\
       & \xi ^{\rho} = 0 \\
       & \xi ^{\phi} = \left( \frac{1}{\zeta ^{+}} - \frac{1}{\zeta ^{-}}\right) \left( \frac{\mathcal{J}^{+}}{\zeta ^{+}} + \frac{\mathcal{J}^{-}}{\zeta ^{-}} \right)^{-1} \Xi (\phi)
  \end{split}
\end{equation}
preserves the fall-off conditions \eqref{24}, up to terms that involve powers of $\delta \mathcal{J}$ higher than the order one, i.e. we ignore the terms of order $\mathcal{O}(\delta \mathcal{J}^{2})$. In the Eq.\eqref{28}, $\Xi (\phi)$ is an arbitrary function of $\phi$. Under the transformation generated by the Killing vector field \eqref{28} the arbitrary functions $\mathcal{J}^{\pm}(\phi)$, which have appeared in the metric, transform as
\begin{equation}\label{29}
  \hat{\delta} _{\xi} \mathcal{J}^{\pm}= \pm \Xi ^{\prime},
\end{equation}
where the prime denotes differentiation with respect to $\phi$. We introduce a modified version of Lie brackets \cite{120}
\begin{equation}\label{30}
  \left[ \xi _{1} , \xi _{2} \right] = \pounds _{\xi _{1}} \xi _{2} - \hat{\delta} _{\xi _{1}} \xi _{2} + \hat{\delta} _{\xi _{2}} \xi _{1},
\end{equation}
so that the algebra of the Killing vector fields to be close. In the equation \eqref{30}, $\hat{\delta} _{\xi _{1}} \xi _{2}$ denotes the change induced in $\xi _{2}$ due to the variation of metric $\delta  _{\xi _{1}} g_{\mu \nu} = \pounds _{\xi _{1}} g_{\mu \nu}$ \cite{6}. Thus, we have
\begin{equation}\label{31}
   \left[ \xi _{1} , \xi _{2} \right] = 0.
\end{equation}
Therefore, the Killing vectors $ \xi _{1} = \xi (\Xi _{1})$ and $ \xi _{2} = \xi (\Xi _{2})$ commute. The relation between metric tensor and dreibein is given by $g_{\mu \nu } = \eta _{a b} e^{a} _{\hspace{1.5 mm} \mu} e^{b} _{\hspace{1.5 mm} \nu}$, so we conclude this section by writing down dreibein correspond to the line-element \eqref{24}
\begin{equation}\label{32}
  \begin{split}
     e^{0} = & - \frac{1}{2} \left[ 2 - \frac{l \rho}{2} \left( f_{+} \zeta ^{+} + f_{-} \zeta ^{-} \right) \right] dv + \frac{l}{2} d \rho \\
       & + \frac{1}{2} \left[ - \left( \frac{\mathcal{J}^{+}}{\zeta ^{+}} - \frac{\mathcal{J}^{-}}{\zeta ^{-}} \right) + \frac{l \rho}{2} \left( f_{+} \mathcal{J}^{+} - f_{-} \mathcal{J}^{-} \right) \right] d \phi \\
     e^{1} = & \frac{l}{2} \left( \zeta ^{+} - \zeta ^{-} \right) dv + \frac{l}{2} \left[ \left(\mathcal{J}^{+}+\mathcal{J}^{-} \right) - \frac{\rho}{l} \left( \frac{\mathcal{J}^{+}}{\zeta ^{+}} + \frac{\mathcal{J}^{-}}{\zeta ^{-}} \right)  \right] d \phi \\
     e^{2} = & - \frac{1}{2} \left[ 2 + \frac{l \rho}{2} \left( f_{+} \zeta ^{+} + f_{-} \zeta ^{-} \right) \right] dv - \frac{l}{2} d \rho \\
       & - \frac{1}{2} \left[ \left( \frac{\mathcal{J}^{+}}{\zeta ^{+}} - \frac{\mathcal{J}^{-}}{\zeta ^{-}} \right) + \frac{l \rho}{2} \left( f_{+} \mathcal{J}^{+} - f_{-} \mathcal{J}^{-} \right) \right] d \phi . \\
  \end{split}
\end{equation}
\section{Apply to the generalized minimal massive gravity}
Generalized minimal massive gravity (GMMG) is an example of the Chern-Simons-like theories of gravity \cite{5'}. This model is realized
by adding the CS deformation term, the higher derivative deformation term, and an extra term
to pure Einstein gravity with a negative cosmological constant. In \cite{5'} it is discussed that this theory is free of negative-energy bulk modes, and also avoids the aforementioned ``bulk-boundary unitarity clash''. By a Hamiltonian analysis one can show that the GMMG model has
no the Boulware-Deser ghosts and this model propagate only two physical modes. In the GMMG, there are four flavours of one-form, $a^{r}= \{ e, \omega , h, f \}$ and the non-zero components of the flavour metric and the flavour tensor are
\begin{equation}\label{33}
\begin{split}
     & \tilde{g}_{e \omega}=-\sigma, \hspace{1 cm} \tilde{g}_{e h}=1, \hspace{1 cm} \tilde{g}_{\omega f}=-\frac{1}{m^{2}}, \hspace{1 cm} \tilde{g}_{\omega \omega}=\frac{1}{\mu}, \\
     & \tilde{f}_{e \omega \omega}=-\sigma, \hspace{1 cm} \tilde{f}_{e h \omega}=1, \hspace{1 cm} \tilde{f}_{f \omega \omega}=-\frac{1}{m^{2}}, \hspace{1 cm} \tilde{f}_{\omega \omega \omega}=\frac{1}{\mu},\\
     & \tilde{f} _{eff}= -\frac{1}{m^{2}}, \hspace{1 cm} \tilde{f}_{eee}=\Lambda_{0},\hspace{1 cm} \tilde{f}_{ehh}= \alpha .
\end{split}
\end{equation}
where $\sigma$, $\Lambda _{0}$, $\mu$, $m$ and $\alpha$ are a sign, cosmological parameter with dimension of mass squared, mass parameter of the Lorentz Chern-Simons term, mass parameter of the new massive gravity term and a dimensionless parameter, respectively. In this case, the equations of motion \eqref{7} reduced to the following equations
\begin{equation}\label{34}
   - \sigma R (\omega) + \frac{\Lambda _{0}}{2} e \times e + D(\omega) h - \frac{1}{2 m^{2}} f \times f  + \frac{\alpha}{2} h \times h =0  ,
\end{equation}
\begin{equation}\label{35}
  - \sigma T(\omega) + \frac{1}{\mu} R(\omega) - \frac{1}{m^{2}} D (\omega) f + e \times h =0 ,
\end{equation}
\begin{equation}\label{36}
    R(\omega) + e \times f =0 ,
\end{equation}
\begin{equation}\label{37}
   T(\omega) + \alpha e \times h =0 .
\end{equation}
Dreibein \eqref{32} solve the equations of motion \eqref{34}-\eqref{37} when the following equations are satisfied \cite{140}
\begin{equation}\label{38}
  f^{a}=Fe^{a}, \hspace{1 cm} h^{a}=He^{a},
\end{equation}
\begin{equation}\label{39}
  \frac{\sigma}{ l^{2}} - \alpha (1 + \sigma \alpha ) H ^{2} + \Lambda _{0} - \frac{F^{2}}{ m^{2}}=0,
\end{equation}
\begin{equation}\label{40}
  - \frac{1}{\mu l^{2}} + 2 (1 + \sigma \alpha ) H + \frac{2 \alpha}{m^{2}} F H + \frac{\alpha ^{2}}{\mu} H^{2}=0,
\end{equation}
\begin{equation}\label{41}
  - F + \mu (1 + \sigma \alpha ) H + \frac{\mu \alpha}{m^{2}} FH=0,
\end{equation}
where $F$ and $H$ are constant parameters. It should be noted that one can decompose the spin-connection in two independent parts $\omega = \Omega +\kappa$, where $\Omega$ is the torsion-free part which is known as the Riemannian spin-connection and $\kappa$ is the contorsion 1-form. It is easy to check that (using Eq.\eqref{37}) the contorsion 1-form for this case is given as $\kappa= \alpha h$.\\
By using equations \eqref{38}-\eqref{41} and $\omega = \Omega + \alpha h$, one can simplify Eq.\eqref{22} in the context of GMMG as
\begin{equation}\label{42}
\begin{split}
   \hat{\delta} Q(\xi) = \frac{1}{8 \pi G} \int _{\Sigma} \{ & - \left( \sigma + \frac{\alpha H}{\mu} + \frac{F}{m^{2}} \right) \left( (i_{\xi} \Omega - \chi _{\xi} ) \cdot \hat{\delta} e + i_{\xi} e \cdot \hat{\delta} \Omega \right) \\
     &  + \frac{1}{\mu} \left( (i_{\xi} \Omega - \chi _{\xi} ) \cdot \hat{\delta} \Omega + \frac{1}{l^{2}} i_{\xi} e \cdot \hat{\delta} e \right) \}.
\end{split}
\end{equation}
Parameters appeared in the Eq.\eqref{42} ($F$ and $H$), satisfy equations \eqref{39}-\eqref{41}. In the one hand since the torsion free spin-connection is given as
\begin{equation}\label{43}
  \Omega ^{a} _{ \hspace{1.5 mm}\mu} = \frac{1}{2} \varepsilon^{a b c} e _{b} ^{ \hspace{1.5 mm} \alpha} \overset{\bullet}{\nabla} _{\mu} e_{c \alpha}
\end{equation}
where $\overset{\bullet}{\nabla}$ denotes covariant derivative with respect to the Levi-Civita connection, then by substituting Eq.\eqref{32} into Eq.\eqref{43} we find that
\begin{equation}\label{44}
  \begin{split}
     \Omega ^{0} = & - \frac{1}{4} \left( \zeta^{+} - \zeta^{-} \right) \rho d v + \frac{1}{2l} \left[ \left( \frac{\mathcal{J}^{+}}{\zeta ^{+}} + \frac{\mathcal{J}^{-}}{\zeta ^{-}} \right) - \frac{l \rho}{2} \left( f_{+} \mathcal{J}^{+} + f_{-} \mathcal{J}^{-} \right) \right] d \phi \\
      \Omega ^{1} = & - \frac{1}{2} \left[ \left( \zeta^{+} + \zeta^{-} \right) - \frac{2 \rho}{l}  \right] dv - \frac{1}{2} \left[ \left( 1 - \frac{\rho}{l \zeta^{+}}\right) \mathcal{J}^{+} - \left( 1 - \frac{\rho}{l \zeta^{-}}\right) \mathcal{J}^{-} \right] d \phi \\
      \Omega ^{2} = & \frac{1}{4} \left( \zeta^{+} - \zeta^{-} \right) \rho d v + \frac{1}{2l} \left[ \left( \frac{\mathcal{J}^{+}}{\zeta ^{+}} + \frac{\mathcal{J}^{-}}{\zeta ^{-}} \right) + \frac{l \rho}{2} \left( f_{+} \mathcal{J}^{+} + f_{-} \mathcal{J}^{-} \right) \right] d \phi .
  \end{split}
\end{equation}
On the other hand, by demanding that the Lie-Lorentz derivative of $e^{a}$ becomes zero explicitly when $\xi$ is a Killing vector field, we find the following expression for $\chi_{\xi}$ \cite{150,3}
\begin{equation}\label{45}
  \chi _{\xi} ^{a} = i_{\xi} \omega ^{a} + \frac{1}{2} \varepsilon ^{a}_{\hspace{1.5 mm} bc} e^{\nu b} (i_{\xi} T^{c})_{\nu} + \frac{1}{2} \varepsilon ^{a}_{\hspace{1.5 mm} bc} e^{b \mu} e^{c \nu} \overset{\bullet}{\nabla} _{\mu} \xi _{\nu} .
\end{equation}
It has been shown that this expression can be rewritten as \cite{160}
\begin{equation}\label{46}
  i_{\xi} \Omega - \chi _{\xi} = - \frac{1}{2} \varepsilon ^{a}_{\hspace{1.5 mm} bc} e^{b \mu} e^{c \nu} \overset{\bullet}{\nabla} _{\mu} \xi _{\nu} .
\end{equation}
Thus, using equations \eqref{28}, \eqref{32}, \eqref{44} and \eqref{46}, we find that
\begin{equation}\label{47}
\begin{split}
   (i_{\xi} \Omega - \chi _{\xi} ) \cdot \hat{\delta} e + i_{\xi} e \cdot \hat{\delta} \Omega = & - \frac{l}{2} \left(  \Xi \hat{\delta} \mathcal{J}^{+} + \Xi \hat{\delta} \mathcal{J}^{-} \right) d \phi + \mathcal{O}( \hat{\delta} \mathcal{J}^{2}), \\
    (i_{\xi} \Omega - \chi _{\xi} ) \cdot \hat{\delta} \Omega + \frac{1}{l^{2}} i_{\xi} e \cdot \hat{\delta} e = &  \frac{1}{2} \left(  \Xi \hat{\delta} \mathcal{J}^{+} - \Xi \hat{\delta} \mathcal{J}^{-} \right) d \phi + \mathcal{O}( \hat{\delta} \mathcal{J}^{2}).
\end{split}
\end{equation}
By substituting Eq.\eqref{47} into Eq.\eqref{42}, then by taking an integration over the one-parameter path on the solution space, we obtain
\begin{equation}\label{48}
  Q(\xi)= Q(\tau ^{+}) + Q(\tau ^{-})
\end{equation}
where $\tau ^{\pm} = \pm \Xi (\phi)$ and $Q(\tau ^{\pm})$ are given as
\begin{equation}\label{49}
  Q(\tau ^{\pm}) = \pm \frac{k}{4 \pi} \left( \sigma \pm \frac{1}{\mu l}+ \frac{\alpha H}{\mu} + \frac{F}{m^{2}} \right) \int_{0}^{2 \pi} \tau ^{\pm} (\phi) \mathcal{J}^{\pm} (\phi) d \phi .
\end{equation}
In the equation \eqref{49} we set $k=l/(4G)$. The algebra of conserved charges can be written as \cite{170}
\begin{equation}\label{50}
  \left\{ Q(\xi _{1}) , Q(\xi _{2}) \right\} = Q \left(  \left[ \xi _{1} , \xi _{2} \right] \right) + \mathcal{C} \left( \xi _{1} , \xi _{2} \right)
\end{equation}
where $\mathcal{C} \left( \xi _{1} , \xi _{2} \right)$ is central extension term. Also, the left hand side of the equation \eqref{50} can be defined by
\begin{equation}\label{51}
  \left\{ Q(\xi _{1}) , Q(\xi _{2}) \right\}= \hat{\delta} _{\xi _{2}} Q(\xi _{1}).
\end{equation}
 Due to the Eq.\eqref{31} one can deduce that $\hat{\delta} _{\xi _{2}} Q(\xi _{1}) = \mathcal{C} \left( \xi _{1} , \xi _{2} \right)$. By varying Eq.\eqref{49} with respect to the dynamical fields so that the variation is generated by a Killing vector, we have
\begin{equation}\label{52}
  \begin{split}
     \hat{\delta} _{\tau _{2} ^{\pm}} Q(\tau _{1} ^{\pm}) = & \pm \frac{k}{8 \pi} \left( \sigma \pm \frac{1}{\mu l}+ \frac{\alpha H}{\mu} + \frac{F}{m^{2}} \right) \int_{0}^{2 \pi} \Xi _{12}(\phi) d \phi , \\
       \hat{\delta} _{\tau _{2} ^{\pm}} Q(\tau _{1} ^{\mp}) = & 0 ,
  \end{split}
\end{equation}
where
\begin{equation}\label{53}
  \Xi _{12} = \Xi _{1} \Xi_{2}^{\prime}-\Xi _{2} \Xi_{1}^{\prime}.
\end{equation}
By setting $\tau ^{\pm} = \pm \Xi (\phi) = \pm e^{in\phi}$, one can expand $Q (\tau ^{\pm})$ in Fourier modes
\begin{equation}\label{54}
  J_{n}^{\pm} = \frac{k}{4 \pi} \left( \sigma \pm \frac{1}{\mu l}+ \frac{\alpha H}{\mu} + \frac{F}{m^{2}} \right) \int_{0}^{2 \pi} e^{in\phi} \mathcal{J}^{\pm} (\phi) d \phi .
\end{equation}
Also, by substituting $\Xi _{1} = e^{in\phi}$, $\Xi _{2} = e^{im\phi}$ into Eq.\eqref{52} and replacement of Dirac brackets
by commutators $ \left\{ , \right\} \rightarrow i \left[ ,\right]$,  we have
\begin{equation}\label{55}
\begin{split}
   \left[ J_{n}^{\pm} , J_{m}^{\pm} \right]=& \mp \frac{k}{2} \left( \sigma \pm \frac{1}{\mu l}+ \frac{\alpha H}{\mu} + \frac{F}{m^{2}} \right) n \delta _{m+n,0}, \\
     \left[ J_{n}^{\pm} , J_{m}^{\mp} \right] = & 0.
\end{split}
\end{equation}
Similar to the near horizon symmetry algebra of the black flower solutions of the Einstein gravity in the presence of negative cosmological constant, the above algebra consists
of two $U(1)$ current algebras, but instead with levels $\pm\frac{k}{2}$, here the level of algebra is given by $\mp \frac{k}{2} \left( \sigma \pm \frac{1}{\mu l}+ \frac{\alpha H}{\mu} + \frac{F}{m^{2}} \right)$.\\
One can change the basis according to following definitions
\begin{equation}\label{56}
\begin{split}
   X_{n} = & \frac{1}{\sqrt{2 u_{+}}} J_{n}^{+} - \frac{i}{\sqrt{2 u_{-}}} J_{n}^{-} \hspace{0.5 cm}  \text{for} \hspace{0.3 cm} n \in \mathbb{Z} \\
   P_{n} = & \frac{i}{n\sqrt{2 u_{+}}} J_{-n}^{+} - \frac{1}{n\sqrt{2 u_{-}}} J_{-n}^{-} \hspace{0.5 cm}  \text{for} \hspace{0.3 cm} n \neq 0 \\
    P_{0} = & J_{0}^{+} + J_{0}^{-} \hspace{0.5 cm}  \text{for} \hspace{0.3 cm} n = 0,
\end{split}
\end{equation}
where
\begin{equation}\label{57}
  u_{\pm} =\mp \frac{k}{2} \left( \sigma \pm \frac{1}{\mu l}+ \frac{\alpha H}{\mu} + \frac{F}{m^{2}} \right).
\end{equation}
By using  the above equations, the algebra \eqref{55}, takes following form
\begin{equation}\label{58}
  \left[ X_{n} , X_{m} \right] = \left[ P_{n} , P_{m} \right] = \left[ X_{0} , P_{n} \right] = \left[ P_{0} , X_{n} \right] =0
\end{equation}
\begin{equation}\label{59}
  \left[ X_{n} , P_{m} \right] = i \delta _{nm} \hspace{0.5 cm} \text{for} \hspace{0.3 cm} n,m \neq 0.
\end{equation}
It is clear that $ X_{0}$ and $ P_{0}$ are the two Casimirs and Eq.\eqref{59} is the Heisenberg algebra.
It is interesting that, for the GMMG model also we obtain the Heisenberg algebra as the near horizon symmetry algebra of the black flower solutions.
By comparing the definition of $P_{0}$ and Eq.\eqref{48}, one can deduce that $P_{0}$ is just the Hamiltonian, i.e. $H \equiv P_{0}$.\\
By setting $\sigma =-1$, $\mu \rightarrow \infty$ and $m^{2} \rightarrow \infty$, the results of this work, namely Eq.\eqref{49}, Eq.\eqref{54} and Eq.\eqref{55}, which we obtained for the Chern-Simons-like theories of gravity, reduced to the results of the Einstein gravity with negative cosmological constant case which have obtained in \cite{9} by a different way.
\section{Soft hair and the soft hairy black hole entropy}
We know that the Hamiltonian $H \equiv P_{0}$ give us the dynamics of the system near the horizon.
Let us consider all vacuum descendants \cite{9}
\begin{equation}\label{60}
  | \psi (q) \rangle = N(q) \prod _{i=1} ^{N^{+}} \left( J_{-n_{i}^{+}}^{+} \right)^{m_{i}^{+}} \prod _{i=1} ^{N^{-}} \left( J_{-n_{i}^{-}}^{-} \right)^{m_{i}^{-}} | 0 \rangle
\end{equation}
where $q$ is a set of arbitrary non-negative integer quantum numbers $N^{\pm}$, $n_{i}^{\pm}$ and $m_{i}^{\pm}$. Also, $N(q)$ is a normalization constant such that $\langle \psi (q) | \psi (q) \rangle = 1$. The Hamiltonian $H \equiv P_{0}=J_{0}^{+} + J_{0}^{-}$ commutes with all generators $J_{n}^{\pm}$, so the energy of all states are the same. The energy of the vacuum state is given by the following eigenvalue equation
\begin{equation}\label{61}
  H  | 0 \rangle = E_{\text{vac}}  | 0 \rangle .
\end{equation}
Also, for all descendants, we have
\begin{equation}\label{62}
  E_{\psi} = \langle \psi (q) | H  | \psi (q) \rangle .
\end{equation}
Due to the mentioned property of the Hamiltonian, we find that all descendants of the vacuum have the same energy as the vacuum,
\begin{equation}\label{63}
  E_{\psi} = E_{\text{vac}},
\end{equation}
in other words, they are soft hairs in the sense of being zero-energy excitations \cite{18,9}.\\
For the case of the BTZ black hole, we have
\begin{equation}\label{64}
  \mathcal{J}^{\pm}= \frac{1}{l} \left( r_{+} \pm r_{-} \right), \hspace{0.7 cm} \zeta^{\pm} = -\frac{r_{+}^{2} - r_{-}^{2}}{l^{2} r_{+}},
\end{equation}
where $r_{-}$ and $r_{+}$ are inner and outer horizon radiuses of the BTZ black hole \cite{19,20}. By substituting Eq.\eqref{64} into Eq.\eqref{54}, we find the eigenvalues of $ J_{n}^{\pm}$ as follows:
\begin{equation}\label{65}
  J_{n}^{\pm} = \frac{1}{8G} \left( \sigma \pm \frac{1}{\mu l}+ \frac{\alpha H}{\mu} + \frac{F}{m^{2}} \right) \left( r_{+} \pm r_{-} \right) \delta _{n,0} .
\end{equation}
The entropy of a soft hairy black hole is related to the zero mode charges $J_{0}^{\pm}$ by the following formula \cite{21,22,23,24,9}
\begin{equation}\label{66}
  S=2\pi \left( J_{0}^{+} + J_{0}^{-} \right).
\end{equation}
Hence, by substituting Eq.\eqref{65} into Eq.\eqref{66}, we find the entropy of the BTZ black hole solution of GMMG as
\begin{equation}\label{67}
  S= - \frac{ \pi}{2G} \left\{ \left( \sigma + \frac{\alpha H}{\mu} + \frac{F}{m^{2}} \right) r_{+} + \frac{r_{-}}{\mu l} \right\}
\end{equation}
which is exactly matched with the results of the paper \cite{25}. As we know, $J_{0}^{+} + J_{0}^{-}= P_{0}$  is one of two Casimirs of the algebra, i.e. $P_{0}$ is a constant of motion. Therefore, one expects that the zero mode eigenvalue of $P_{0}$ should be corresponds to a conserved charge of considered spacetime. We have shown that entropy is the intended conserved charge in the context of GMMG, as the pure-gravity case.
\section{Conclusion}
Our aim in this paper was that study the near horizon symmetry algebra of the black hole solutions of the Chern-Simons-like theories of gravity, which are stationary but are not necessarily spherically symmetric. The Lagrangian of such theories are given by Eq.\eqref{1} in the first order formalism. We have tried to find an expression for the quasi-local conserved charges of CSLTG associated with the field dependent Killing vector fields. For this purpose, we have used the concept of total variation \eqref{3} to define an off-shell conserved current \eqref{12}. We took a variation from Eq.\eqref{12} with respect to dynamical fields and then we defined the extended off-shell ADT current \eqref{15}. We  have shown that the extended off-shell ADT current is an extension of the generalized off-shell ADT current, i.e. we have extended the generalized off-shell ADT current such that it becomes conserved for the field dependent Killing vectors and the field dependent asymptotically Killing vectors. So this experssion reduced to the generalized off-shell ADT current \cite{4} when $\xi$ is independent of dynamical fields, i.e, where $\hat{\delta} \xi =0$. Then, we have found extended off-shell ADT conserved charge associated with the field dependent Killing vector field \eqref{21}. Consequently, we have defined the quasi-local conserved charge corresponds to a field dependent Killing vector field as Eq.\eqref{23} which is conserved for the field dependent asymptotically Killing vectors as well. In section 3, we have considered  the extended near horizon geometry which have been proposed in \cite{9}. In the paper \cite{9}, the metric \eqref{24} was introduced as new fall-off conditions for the near horizon of a non-extremal black hole in 3D. This geometry is not spherically symmetric, and generically describes a "black flower" \cite{16}. We have shown that Killing vectors of the form \eqref{28} preserve that fall-off conditions up to terms that involve powers of perturbations of dynamical fields higher than the one. In section 4, we have applied the provided formalism to the generalized minimal massive gravity as an example of the Chern-Simons-like theories of gravity. We have found the conserved charges correspond to the near horizon symmetry of a non-extremal non-spherically symmetric black hole solution of GMMG, see Eq.\eqref{48} and Eq. \eqref{49}. Then, we have obtained the algebra of conserved charges in Fourier modes, see Eq.\eqref{55} or Eq.\eqref{58} and Eq.\eqref{59}.
 It is interesting that, similar to the Einstein gravity in the presence of negative cosmological constant, for the GMMG model also we obtain the Heisenberg algebra as the near horizon symmetry algebra of the black flower solutions. In the section 5, we have summarized the concept of soft hair presented in \cite{9} and we have argued that it is valid in GMMG also. In the another words, since the Hamiltonian is given by  $H \equiv P_{0}$, and  $P_{0}$ is a Casimir of the algebra, so the vacuum state and all descendants of the vacuum have the same energy. So these zero energy excitations on horizon appear as soft hair on the black hole. Then by finding the eigenvalues of $ J_{n}^{\pm}$ for the BTZ black hole, see Eq.\eqref{65}, we have checked that the formula for the entropy of a soft hairy black hole gives us the correct value of the entropy of the BTZ black hole solution of the GMMG. It should be mentioned that, as one expected, by setting $\sigma =-1$, $\mu \rightarrow \infty$ and $m^{2} \rightarrow \infty$, where the GMMG reduce to the Einstein gravity with negative cosmological constant, the results of this paper, namely Eq.\eqref{49}, Eq.\eqref{54} and Eq.\eqref{55}, reduced to the results of \cite{9} which have been obtained by a different way.

\section{Acknowledgments}
M. R. Setare thank Stephane Detournay for helpful comments and discussions.

\end{document}